\documentclass[12pt]{article}
\usepackage{epsf}
\usepackage{amsmath}
\usepackage{graphics}

\renewcommand{\thefootnote}{\#\arabic{footnote}}
\begin{document}

\newcommand{\gtrsim}{ \mathop{}_{\textstyle \sim}^{\textstyle >} }
\newcommand{\lesssim}{ \mathop{}_{\textstyle \sim}^{\textstyle <} }

\renewcommand{\thefootnote}{\fnsymbol{footnote}}
\setcounter{footnote}{0}
\begin{titlepage}

\def\thefootnote{\fnsymbol{footnote}}

\begin{center}

\hfill astro-ph/0305495\\
\hfill May 2003\\

\vskip .5in

{\Large \bf
Cosmic Degeneracy with Dark Energy \\
Equation of State
}

\vskip .45in

{\large
James L. Crooks, James O. Dunn, Paul H. Frampton, \\
Heather R. Norton and Tomo Takahashi
}

\vskip .45in

{\em
Department of Physics and Astronomy,\\
University of North Carolina, Chapel Hill, NC 27599-3255, USA
}

\end{center}

\vskip .4in

\begin{abstract}
  We discuss a degeneracy between the geometry of the universe and the
  dark energy equation of state $w_{\rm X}$ which exists in the power
  spectrum of the cosmic microwave background.  In particular, for the
  case with $w_{\rm X} < -1$, this degeneracy has interesting
  implications to a lower bound on $w_{\rm X}$ from observations.  It
  is also discussed how this degeneracy can be removed using current
  observations of type Ia supernovae.

\end{abstract}
\end{titlepage}

\renewcommand{\thepage}{\arabic{page}}
\setcounter{page}{1}
\renewcommand{\thefootnote}{\#\arabic{footnote}}

\section{Introduction}
\label{sec:introduction}
\setcounter{equation}{0}
\setcounter{footnote}{0}

Almost all observations strongly suggest that the present universe is
dominated by an enigmatic component called dark energy.  The simplest
candidate for dark energy is the cosmological constant; other models
such as quintessence \cite{quintessence}, k-essence \cite{k-essence},
that of stringy origin \cite{Bastero-Gil_et_al_2002,Frampton_2002},
the generalized Chaplygin gas \cite{chaplygin} and Cardassian models
\cite{cardasian} have also been discussed.  It is very important to
consider whether we can differentiate these models of dark energy from
current observations.  Many authors have studied this subject for
specific models or in a phenomenological way.  Phenomenologically, the
dark energy is characterized by its equation of state $w_{\rm X} =
p_{\rm X}/\rho_{\rm X}$ where $\rho_{\rm X}$ and $p_{\rm X}$ are its
energy density and pressure respectively.  For the cosmological
constant, $w_{\rm X}$ is equal to $-1$; however, in other models, it
can possibly deviate from $-1$. Several authors have studied
constraints on $w_{\rm X}$ from observations such as cosmic microwave
background (CMB), large scale structure (LSS) and type Ia Supernovae
(SNeIa) \cite{Bean_et_al_2001,Hannestad_2002,Melchiorri_et_al_2002}.
Since observations of CMB strongly suggest that the universe is flat
\cite{Wang_et_al_2001,WMAP_spergel}, most analyses were done with the
prior of a flat universe to study constraints on $w_{\rm X}$.  However
it is important to notice that almost all works which obtained
constraints on the geometry of the universe from CMB observations
assumed a cosmological constant as the dark energy, i.e., equation of
state $w_{\rm X}=-1$.  It is well-known that the peak location of
the power spectrum of CMB is sensitive to the geometry of the
universe. However, it is also known that the equation of state of dark
energy affects the location of acoustic peaks.  Thus it is interesting
to study the power spectrum of CMB with varying $\Omega_{\rm k}$ and
$w_{\rm X}$ where $\Omega_{\rm k}$ is the energy density of curvature
normalized by the critical energy density.

In this paper, we discuss observational consequences of varying the
geometry of the universe $\Omega_{\rm k}$ and the equation of state of
dark energy $w_{\rm X}$.  Although many models have time-dependent
equation of state and some papers investigate observational
consequences of such cases \cite{time_dep_w}, to study it
phenomenologically, we assume that $w_{\rm X}$ is independent of time
throughout this paper. By studying the angular power spectrum of the
CMB varying $\Omega_{\rm k}$ and $w_{\rm X}$, one can find that there
is a degeneracy between $\Omega_{\rm k}$ and $w_{\rm X}$ in CMB.  This
means that we cannot know precisely the geometry of the universe
without knowledge of the equation of state of the dark energy from
observations of CMB.  This also means that the constraints on $w_{\rm
X}$ obtained so far which assume a flat universe is changed if we also
consider the case with a non-flat universe\footnote{
In \cite{Aurich_Steiner}, the authors discussed a $(w_{\rm X},
\Omega_{\rm X}, \Omega_{\rm c})$-degeneracy in an open universe.
}.  In most analyses so far, it is assumed the equation of state as
$w_{\rm X} \ge -1$ since the case with $w_{\rm X} < -1$ violates the
weak energy condition.  However, some models which predict $w_{\rm X}
< -1$ have been proposed \cite{Frampton_2002} and some authors discuss
constraints on $w_{\rm X}$ for such cases
\cite{Hannestad_2002,Melchiorri_et_al_2002}.  Thus, in this paper, we
discuss the case with $w_{\rm X} \le -1$ in which the degeneracy
occurs in a closed universe.  It is interesting to notice that the
degeneracy in the region where $w_{\rm X}< -1$ affects a observational
lower bound on $w_{\rm X}$. Compared to the one obtained assuming a
flat universe, a lower bound on $w_{\rm X}$ becomes less stringent.

The organization of this paper is as follows.  In the next section, we
examine the degeneracy between $\Omega_{\rm k}$ and $w_{\rm X}$ in
CMB.  Then, in section 3, we study implications to this degeneracy
from the observation of type Ia supernovae.  The last section is
devoted to summary.

\section{Degeneracy between $\Omega_{\rm k}$ and $w_{\rm X}$ in CMB}
Observations of the CMB can measure cosmological parameters very
precisely as seen from the recent WMAP result \cite{WMAP_spergel}.
However, it is known that there exist some degeneracies in certain
sets of cosmological parameters which cannot be removed from the
observations of the CMB alone
\cite{Zaldarriaga_et_al_1997,Efstathiou_Bond_1998,Heuy_et_al_1999}. One
such example is a degeneracy between $\Omega_{\rm m}$ and $h$ where
$h$ is the Hubble parameter (in units of 100 km/s/Mpc).  Although we
cannot break this degeneracy with observations of CMB, the Hubble
parameter can be independently determined by other observations such
as from the Hubble Space Telescope (HST) \cite{HST}.  Once $h$ is
determined independently of CMB observations, the degeneracy is
lifted.

\begin{figure}
    \centerline{\epsfxsize=0.7\textwidth\epsfbox{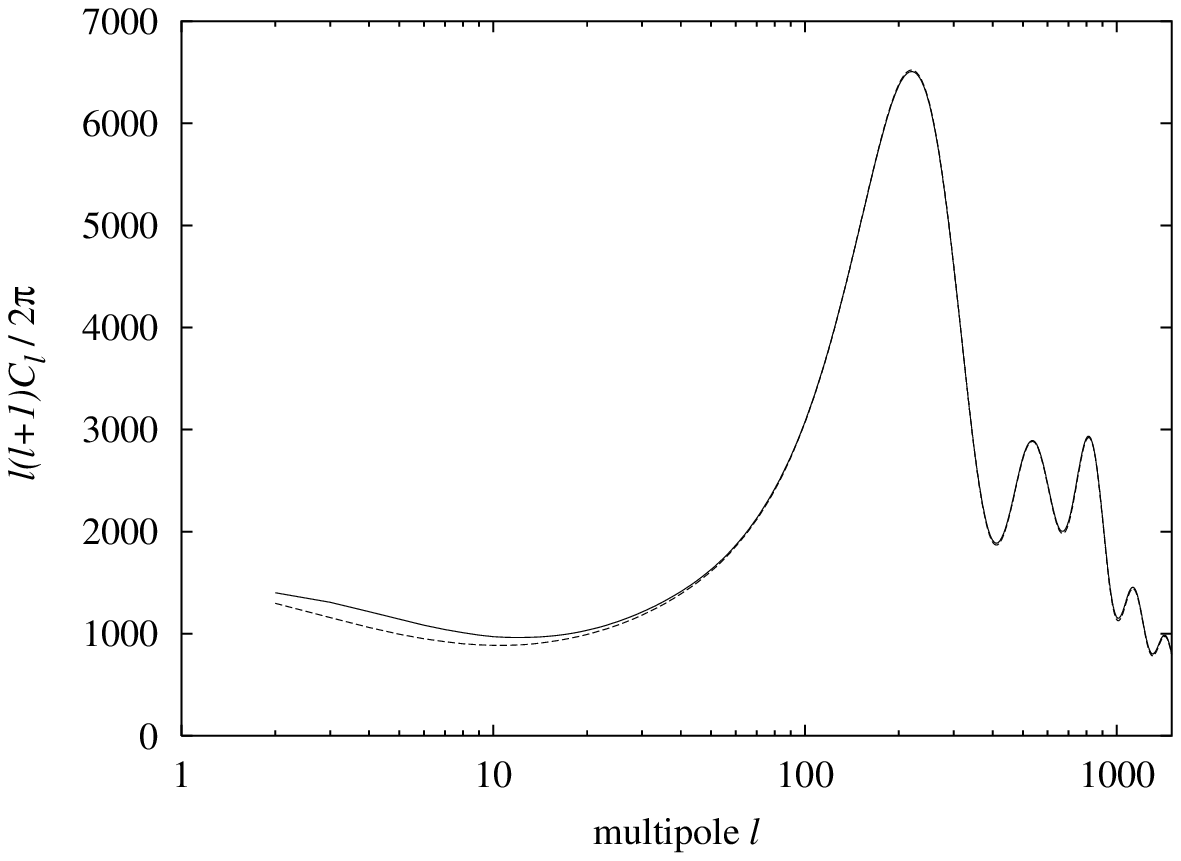}} \caption{The
   angular power spectra of CMB for $\Omega_{\rm k}=0, w=-1$ (solid
   line) and for $\Omega_{\rm k}=-0.05, w_{\rm X}=-5$ (dashed
   line). The other cosmological parameters are taken to be
   $\Omega_{\rm m}h^2=0.14, \Omega_{\rm b}h^2=0.024, h=0.72$, $n=0.99$
   and $\tau =0.166$. For most of the multipole $l$, the dashed and
   solid lines are almost indistinguishable.}
   \label{fig:degeneracy_CMB}
\end{figure}

Here we discuss another degeneracy. As we will show in the following,
there is a degeneracy between $\Omega_{\rm k}$ and $w_{\rm X}$.  To
illustrate this degeneracy, we plot the CMB angular power spectrum
$C_l$ for the cases with $\Omega_{\rm k} = 0, w_{\rm X} = -1$ and with
$\Omega_{\rm k} = -0.05, w_{\rm X} = -5$ in Fig.\
\ref{fig:degeneracy_CMB}. To calculate the angular power spectrum
$C_l$, we used a modified version of CMBFAST \cite{CMBFAST}, and in
the present analysis, we consider constant $w_{\rm X}$ and included
fluctuation of dark energy according to \cite{Hu_1998}.  Here we took
the sound speed in the rest frame of the dark energy as $c_{\rm
eff}=1$.  In this figure, other cosmological parameters are fixed as
$\Omega_{\rm m} = \Omega_{\rm b} + \Omega_{\rm c}=0.27, \Omega_{\rm
b}h^2=0.024$ and $n=0.99$ and $\tau =0.166$ which correspond to the
best-fit values for the power law $\Lambda$CDM model from WMAP results
\cite{WMAP_spergel}, where $\Omega_{\rm m}, \Omega_{\rm b}$ and
$\Omega_{\rm c}$ are the energy density of matter, baryon and cold
dark matter (CDM) respectively, $n$ is the initial spectral index and
$\tau$ is the optical depth of reionization.  From this figure, we can
see the degeneracy between $\Omega_{\rm k}$ and $w_{\rm X}$.  Since
the current constraint on $w_{\rm X}$ assuming the flat universe is
$-1.38 < w_{\rm X} < -0.82$ (at 95 \% C.L.)
\cite{Melchiorri_et_al_2002}, $w_{\rm X}=-5$ is completely disfavored
in the flat universe.  However if we consider such a value in the
non-flat universe, the resulting angular power spectrum can become
almost the same as the best-fit model.  When we take the equation of
state as $w_{\rm X} < -1$, effects of dark energy becomes significant
only at late times, thus the structure of the acoustic peaks is mostly
determined by the energy density of matter (i.e., $\Omega_{\rm c}$ and
$\Omega_{\rm b}$). Therefore the shape of the acoustic peaks is almost
the same even if we change the values of $\Omega_{\rm k}$ and $w_{\rm
X}$. However the location of the acoustic peaks is affected
by $\Omega_{\rm k}$ and $w_{\rm X}$. In a closed universe
(i.e. $\Omega_{\rm k} <0$), the peak location is shifted to lower
multipole. On the other hand, by decreasing $w_{\rm X}$, it is shifted
to higher multipole.  If we take the values of $\Omega_{\rm k}$ and
$w_{\rm X}$ to cancel these effects, we can have almost the same
angular power spectrum for different sets of these parameters.  Notice
that, since the angular power spectrum at low multipole region is
affected by the integrated Sachs-Wolfe (ISW) effect, the different
sets of $\Omega_{\rm k}$ and $w_{\rm X}$ may give different structure
at low multipoles even if they have indistinguishable structure of
acoustic peaks.  However, also notice that even in such a region, it
may be difficult to distinguish them due to the cosmic variance.

\begin{figure}
  \centerline{\epsfxsize=1\textwidth
\epsfbox{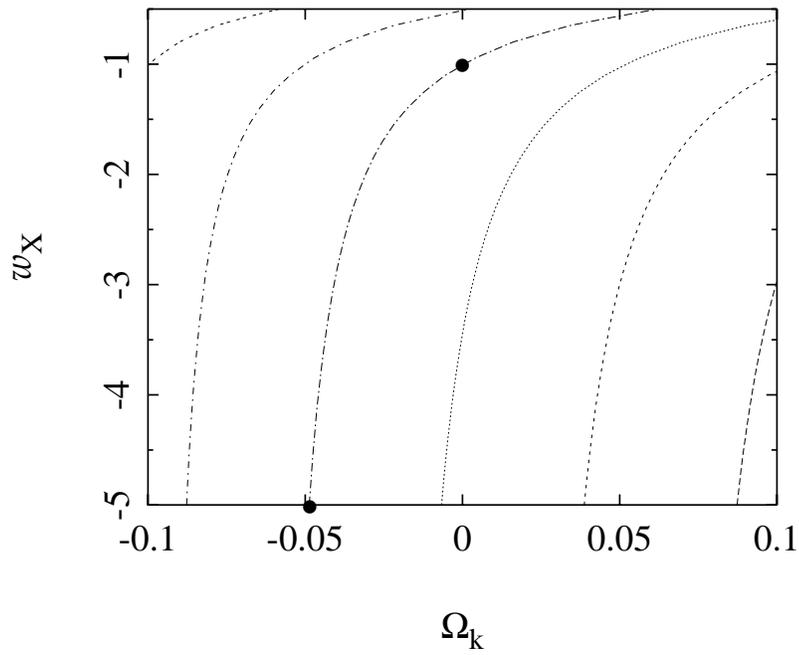}}
   \caption{Contours of constant $d_A$ are shown. The points in this
figure indicate the parameters which are used in Fig.\
\ref{fig:degeneracy_CMB}.}
    \label{fig:d_A}
\end{figure}

Changing $\Omega_{\rm k}$ and $w_{\rm X}$, the location of the peaks
can be shifted because the peak position is mostly determined by the
angular diameter distance to the last scattering surface which can be
affected by $\Omega_{\rm k}$ and $w_{\rm X}$.  The location of the
acoustic peak is inversely proportional to the angular diameter
distance
\footnote{
It also depends on the sound horizon at the surface of last
scattering. However, since we assume that the equation of state is
constant here, the dark energy affects the CMB only at late time,
i.e., the sound horizon at the surface of last scattering is not
affected by changing $w_{\rm X}$.  Thus we do not consider the effect
of the sound horizon on the location of the peaks here.  }
\cite{Hu_Sugiyama,Frampton_et_al_1998}
\begin{eqnarray}
    l_{\rm peak} \propto \frac{1}{d_{\rm A}(z_{\rm rec})}
\end{eqnarray}
where $d_{\rm A}$ is given by
\begin{eqnarray}
d_A(z) &=& \frac{1}{(1+z)|\Omega_{\rm k}|^{1/2}H_0} \\ \notag
&\times &{\rm sinn}
\left\{ |\Omega_{\rm k}|^{1/2}
\int^z_0 dz'
\left[ \Omega_{\rm r}(1+z')^4 +\Omega_{\rm m} (1+z')^3
+ \Omega_{\rm k}(1+z')^2 + \Omega_{\rm X}(1+z')^{3(1+w)} \right]^{-1/2}
\right\}.
\label{eq:d_A}
\end{eqnarray}
Here ``${\rm sinn}$'' is defined as ``$\sin$'' for $\Omega_{\rm k} <
0$, ``1'' for $\Omega_{\rm k} = 0$ and ``$\sinh$'' for $\Omega_{\rm k}
> 0$.  In Fig.\ \ref{fig:d_A}, we plot the angular diameter distance
in the $\Omega_{\rm k}-w_{\rm X}$ plane. The points shown in the
figure indicate the same parameters which are used in Fig.\
\ref{fig:degeneracy_CMB}.  We can clearly see that the parameter sets
which give the same angular diameter distance also give almost the
same peak position.

For intuitive understanding, we consider the effects of the equation
of state and the geometry of the universe with an example.  Now we
compare the case with $w_{\rm X}=-1$ and $w_{\rm X}=-2$.  If $w_{\rm
X}=-2$, the energy density of the dark energy $\rho_{\rm X}$ grows
proportional to $a^3$.  When the present energy density of the dark
energy is fixed as $\Omega_{\rm X}=0.7$, this means that the total
energy density of the universe in the past is smaller than that for
the case with $w_{\rm X}=-1$ since the energy density of dark energy
grows faster than for the $w_{\rm X}=-1$ case.  In other words, the
expansion of the universe is slower than that for the case with
$w_{\rm X}=-1$. This means that the distance to the last scattering
surface becomes larger. Thus, if we take $w_{\rm X}$ more negative,
the locations of peaks are shifted to higher multipoles.  On the other
hand, in a closed universe (i.e., $\Omega_{\rm k} < 0$), the locations
of the acoustic peaks are shifted to lower multipoles compared to the
case with a flat universe because of the geodesic effects on the
trajectories of photons.  So, if we consider the dark energy with
super-negative $w_{\rm X}$ (i.e., $w_{\rm X} < -1$) in a closed
universe, these effects can cancel each other, and the resulting
location of the peaks becomes the same as that of the flat universe
with a cosmological constant.  Conversely, when we consider dark
energy with $w_{\rm X}>-1$ in an open universe, this kind of
cancellation can also happen. But in this case, the dark energy
affects the history of the universe from earlier times than for the
case with $w_{\rm X} < -1$. Thus the degeneracy is milder compared to
the case with a super-negative equation of state, $w_{\rm X} <-1$.

Notice that, if we take the equation of state as $w_{\rm X} \ll -1$,
the dark energy is less relevant to the angular diameter distance.  As
we can see from Eq.\ (\ref{eq:d_A}), for the case with $w_{\rm X} \ll
-1$, the term related to the dark energy becomes less significant in
the integral. This means that the shift of the location of the
acoustic peaks saturates at some point \cite{Caldwell_2002}.  Thus
even if we take very negative values of $w_{\rm X}$, the value of
$\Omega_{\rm k}$ which is need to realize the same peak location as
the $\Lambda$CDM model does not change much.  This means that we
cannot obtain a significant lower bound on $w_{\rm X}$ from CMB
observations alone.

\begin{figure}
    \centerline{\epsfxsize=1\textwidth
\epsfbox{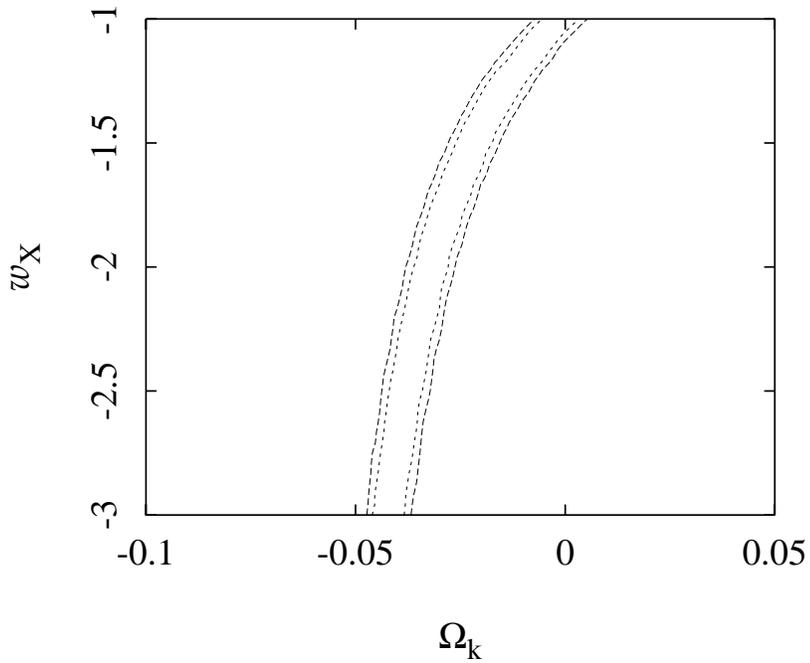}}
   \caption{Contours of constant $\Delta \chi^2$ in the
	$\Omega_{\rm k}-w_{\rm X}$ plane.
	In this figure, we fixed the other cosmological parameters
   	as in Fig.\ \ref{fig:degeneracy_CMB}. The dotted and dashed lines
    	indicate contour for $\Delta \chi^2 = 6.18$ and 11.83  respectively.}
    \label{fig:likelihood_degeneracy}
\end{figure}

To study this degeneracy quantitatively, we show contours of $\Delta
\chi^2$ in the $\Omega_{\rm k}-w_{\rm X}$ plane in Fig.\
\ref{fig:likelihood_degeneracy}.  In this figure, for an illustration,
we fixed the other cosmological parameters except $\Omega_{\rm k}$ and
$w_{\rm X}$ as $\Omega_{\rm m}h^2=0.14, \Omega_{\rm b}h^2=0.024,
h=0.71$, $n=0.99$ and $\tau=0.166$ and the normalization of spectra is
marginalized.  The value of $\chi^2$ is calculated using the code
provided by WMAP team \cite{WMAP_verde,WMAP_hinshaw}.  Here we used
the TT mode data only.  Since locations of the acoustic peaks are
strongly dependent on the angular diameter distance, contours of
constant $\Delta \chi^2$ look similar to $d_A$.  As we can see from
the figure, the allowed region extends to more negative $w_{\rm X}$
region, which indicates that we cannot obtain a significant lower
bound on $w_{\rm X}$ from CMB data alone without specifying the value
of $\Omega_{\rm k}$ as we mentioned above.  This is because effects of
decreasing $w_{\rm X}$ on the peak location saturate at some points.
This also means that we cannot determine the geometry of the universe
without the knowledge of the equation of state of the dark energy.
However if we use other cosmological data such as SNeIa, we can lift
the degeneracy to some extent. We will discuss this issue in the next
section.

\section{Implications from SNe observations}
In this section, we see whether we can break the degeneracy which
exists in the power spectrum of CMB.  It is well-known that confidence
level contours in the $\Omega_{\rm m}-\Omega_{\Lambda}$ plane from CMB
observations lie almost along the flat universe line, while contours
from SNe observations are roughly orthogonal to those from CMB
observations.  So, in the $\Omega_{\rm m}-\Omega_{\Lambda}$ plane,
observations of CMB and SNeIa are complementary. Thus we can expect
that this kind of complementarity exists in the $\Omega_{\rm k}-
w_{\rm X}$ plane.  In the following, we study the constraint with
current SNe data in this plane.

First, we explain some technical details which are needed to consider
constraints from SNeIa observations. Observations of SNeIa give a measured
distance modulus $\mu_0$ which is given by
\begin{equation}
\mu_0 = 5 \log \left(\frac{d_L(z)}{\rm Mpc} \right) + 25,
\label{eq:distance_modulus}
\end{equation}
where $d_L$ is the luminosity distance in units of Mpc. The luminosity
distance is calculated as
\begin{equation}
d_L(z) = \frac{1+z}{|\Omega_{\rm k}|^{1/2}H_0} {\rm sinn}
\left\{ |\Omega_{\rm k}|^{1/2}
\int^z_0 dz'
\left[ \Omega_{\rm m} (1+z')^3 + \Omega_{\rm k}(1+z')^2
	+ \Omega_{\rm X}(1+z')^{3(1+w)} \right]^{-1/2}
\right\}.
\end{equation}
in which definition of ``${\rm sinn}$'' is given in the previous section.

To find constraints from SNeIa observations, we used data from High-z
Supernova Search Team \cite{Reiss_1998} and Supernova Cosmology
Project (SCP) \cite{Perlmutter_1999}. We also included the data from
SNeIa discovered recently \cite{Reiss_2001,Benitez_2002,Blakeslee_2003}.
The published data of High-z Supernova Search Team gives the distance
modulus of each SNeIa defined in Eq.\ (\ref{eq:distance_modulus}),
while the SCP team gives the estimated effective B-band magnitude of
each SN Ia $m_{\rm B}^{\rm eff}$. The effective B-band magnitude
$m_{\rm B}^{\rm eff}$ and the distance modulus $\mu_0$ are related by
\begin{equation}
m_{\rm B}^{\rm eff} = M_{\rm B} + \mu_0.
\end{equation}
where $M_{\rm B}$ is the peak absolute magnitude of a ``standard'' SN Ia
in the B-band.
Wang \cite{Wang_2000} found the relation between the data of SCP team
and High-z SN team
\begin{equation}
M_{\rm B}^{\rm MLCS} \equiv  m_{\rm B}^{\rm MLCS} - \mu_0^{\rm MLCS}
= -19.33 \pm 0.25,
\end{equation}
where all quantities in this equation are estimated with the multi
light curve shape (MLCS) method.

We use this relation to combine data of High-z SN Search team and SCP
team. The data set of High-z SN team and SCP team consist 50 SNe and
42 SNe. In their data set, 18 SNe are the same, thus, including recently
discovered 3 SNe, 95 SNe are used in total in our analysis.

The likelihood for the parameters are determined from a $\chi^2$
statistics,
\begin{equation}
\chi^2 = \sum_i
\frac{[ \mu_0^{\rm th}(z_i; \Omega_{\rm m}, \Omega_{\rm X}, h, w)
- \mu_{0,i}^{\rm exp}]^2}{ \sigma_{\mu 0,i}^2 + \sigma_v^2 }
\end{equation}
where $\mu_0^{\rm th}(z_i; \Omega_{\rm m}, \Omega_{\rm X}, h, w)$ is a
predicted distance modulus for a given cosmological parameters,
$\mu_{0,i}^{\rm exp}$ is the measured distance modulus for each SN,
$\sigma_{\mu 0,i}$ is the measurement error of the distance modulus
and $\sigma_v$ is the velocity dispersion in galaxy redshift in units
of the distance modulus.  For the data of High-z SN team, following
\cite{Reiss_1998}, we adopted $\sigma_v =200~ {\rm km/s}$ and added
2500 km/s in quadrature for high-redshift SN Ia whose redshift were
determined from the broad features in the SN spectrum.  We calculated
the likelihood function as $ \mathcal{L} = \mathcal{L}_0 \exp [
-\chi^2/2 ], $ where $\mathcal{L}_0$ is an arbitrary
normalization constant.

\begin{figure}
\begin{center}
\scalebox{2}{ \includegraphics{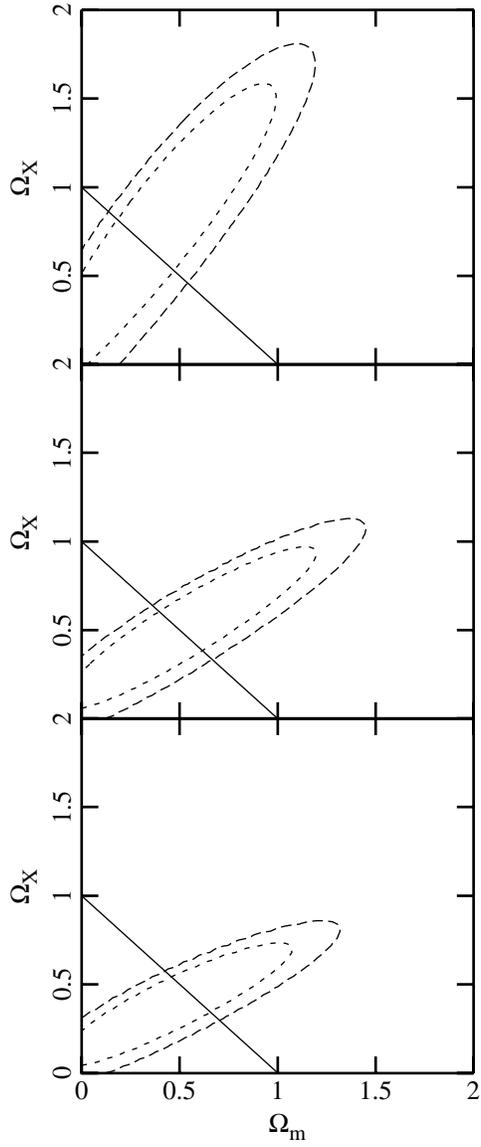}}
\end{center}
  \caption{Contours of constant likelihood in the $\Omega_{\rm
   m}-\Omega_{\rm X}$ plane for $w_{\rm X} = -1$ (top), $w_{\rm X}=-2$
   (middle) and $w_{\rm X}=-3$ (bottom).  The 95.4 \% and 99.7
   \% confidence intervals are shown with dotted and dashed lines
   respectively.}  \label{fig:SN_mxplane}
\end{figure}

First, we show the likelihood contours in the $\Omega_{\rm
m}-\Omega_{\rm X}$ plane in Fig.\ \ref{fig:SN_mxplane}.  In this
figure, the equation of state of dark energy is fixed as $w_{\rm
X}=-1, -2$ and $-3$, respectively. As $w_{\rm X}$ decreases, the
allowed region is shifted to lower values of $\Omega_{\rm X}$. This is
because a lower value of $w_{\rm X}$ can result in a more accelerating
universe for the same value of $\Omega_{\rm X}$. Also notice that,
similarly to the case with the angular diameter distance, the dark
energy is less relevant to the luminosity distance if we take the
equation of state $w_{\rm X}$ at a very negative value. Thus the
likelihood contours for the cases with $w_{\rm X}=-2$ and $-3$ are
less different compared to than those for the $w_{\rm X}=-1$ and
$w_{\rm X}=-2$ cases.

\begin{figure}
\begin{center}
\scalebox{2}{ \includegraphics{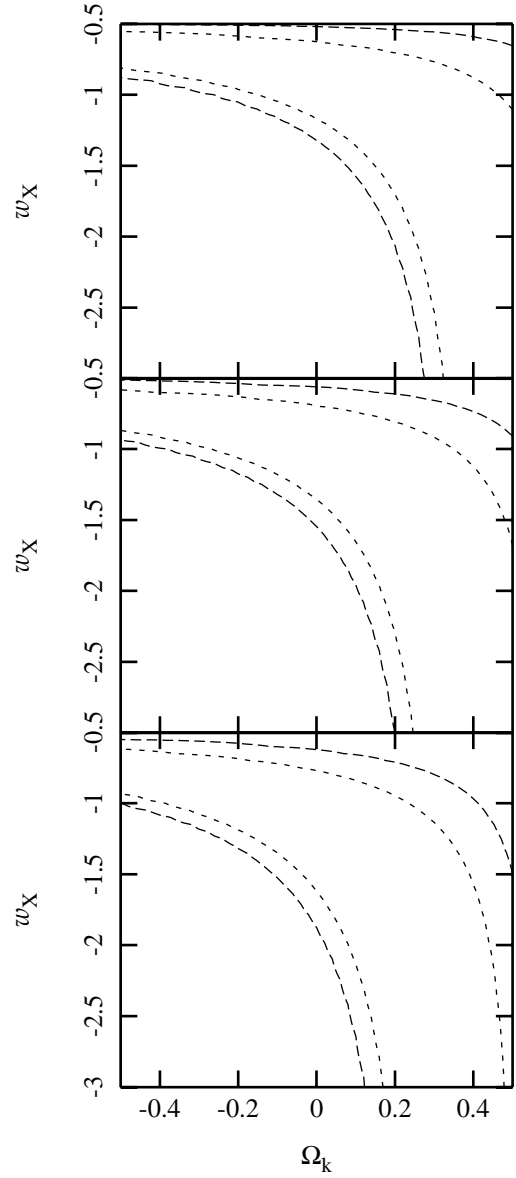}}
\end{center}
  \caption{Contours of constant likelihood in the $\Omega_{\rm
   k}-w_{\rm X}$ plane for $\Omega_{\rm m}=0.25$ (top), $\Omega_{\rm
   m}=0.3$ (middle) and $\Omega_{\rm m}=0.35$ (bottom).  The 95.4 \%
   and 99.7 \% confidence intervals are shown with dotted and dashed
   lines respectively.}
  \label{fig:SN_mwplane}
\end{figure}

In Fig.\ \ref{fig:SN_mwplane}, the likelihood contours in the
$\Omega_{\rm k}-w_{\rm X}$ plane are shown. In this figure, the density
parameter for matter is fixed as $\Omega_{\rm m}=0.25, 0.3$ and 0.35.
As we can see from this figure, allowed regions extends to very
negative values of $w_{\rm X}$. The situation is similar to the case
with CMB which we discussed in the previous section. However, the
allowed parameter space which extends to more negative region in Fig.\
\ref{fig:SN_mwplane} is different from that in the case of CMB (see
Fig.\ \ref{fig:likelihood_degeneracy}).\footnote{
Even though we can see that allowed regions extends to very negative
values of $w_{\rm X}$ as in the case of CMB, this is not an intrinsic
degeneracy. We can have a significant lower bound on $w_{\rm X}$ using
future SNe observations such as SNAP \cite{snap_web}.}
  In this sense, observations of SNe and CMB are complementary each
other in the $\Omega_{\rm k}-w_{\rm X}$ plane too.  Thus we can break
the degeneracy in CMB to some extent with observations of SNeIa.  But
it is important to notice that, even if we can break the degeneracy
between $\Omega_{\rm k}$ and $w_{\rm X}$, the constraint on $w_{\rm
X}$ without assuming the flat universe is less stringent than that
obtained assuming the flat universe. Thus, to impose a more general
lower bound on $w_{\rm X}$ would require a new global fit to all
cosmological parameters with the prior of the flat universe lifted.
Detailed investigation of this issue is beyond the scope of this
paper.

\section{Summary}

We discussed the degeneracy between the geometry of the universe and
the equation of state of dark energy which exists in the CMB power
spectrum.  As is well-known, the geometry of the universe affects the
location of the acoustic peaks. For the open universe, the peak
position is shifted to higher multipole compared to that of the flat
universe, while, for the closed universe, it is shifted to lower
multipoles. It is also known that the equation of state of dark energy
affects the location of the peaks. If we take a smaller value of
$w_{\rm X}$, the peak location is shifted to higher multipoles, and
for a larger value of $w_{\rm X}$, it becomes lower. If we take the
values of $\Omega_{\rm k}$ and $w_{\rm X}$ to cancel these effects, we
can have almost the same angular power spectrum for different sets of
these parameters.  For example, even if we take $w_{\rm X}=-5$, we can
have a power spectrum which has good agreement with CMB data for
models which predict moderate ISW effect at low multipole by choosing
the non-flat universe as $\Omega_{\rm k}=-0.05$.  Since effects of
decreasing $w_{\rm X}$ on the location of the acoustic peaks saturate
at some points, we cannot obtain a significant lower bound on $w_{\rm
X}$ from CMB observations alone if we remove the assumption of the flat
universe.  However, observation of SNeIa can remove this degeneracy to
some extent, as discussed in section 3.  For the degeneracy that we
discussed here, observations of CMB and SNeIa are complementary.  But
it is interesting to notice that, when we relax the assumption of the
flat universe, the constraint on $w_{\rm X}$ becomes less stringent
compared to that assuming a flat universe.  The full investigation of
this issue will be the subject of a future work.

\bigskip
\bigskip

\noindent
{\bf Acknowledgment:} This work was supported in part by the US
Department of Energy under Grant No. DE-FG02-97ER-41036.  We
acknowledge the use of CMBFAST \cite{CMBFAST}.

\pagebreak

\end{document}